\documentclass[fleqn,twoside]{article}
\usepackage[headings]{espcrc2}

\usepackage{amsmath,amssymb}
\usepackage{graphicx}
\usepackage{paralist}
\usepackage{url}
\usepackage{hyperref}
\usepackage{xspace}

\newcommand{\fastjet}{{\sf FastJet}\xspace}
\newcommand{\order}[1]{{\cal{O}}\left(#1\right)}
\newcommand{\rhol}{\ensuremath{\rho_{\cal{L}}}\xspace}
\newcommand{\JD}{\mathrm{JD}\xspace}
\newcommand{\Qa}[1]{\ensuremath{Q_{f=#1}^{w}}\xspace}

\title{Recent progress in defining jets}

\author{Gr\'egory Soyez\address{Brookhaven National Laboratory,
    Building 510, Upton NY 11973, USA}%
  \thanks{Work done under Contract No. DE-AC02-98CH10886 with the
    U.S. Department of Energy.}}
       
\begin{document}

\begin{abstract}
  \vspace{1pc} From dedicated QCD studies to new physics background
  estimation, jets will be everywhere at the LHC. In these
  proceedings, we discuss two important recent series of improvements.
  In the first one, we introduce new algorithms and new
  implementations of previously existing algorithms, in order to cure
  limitations of their predecessors and to satisfy fundamental
  requirements.
  In the second part, we show that it is of prime importance to
  carefully choose the jet definition --- algorithm and parameters ---
  to optimise kinematic reconstructions at the LHC. Noticeably, we
  show that while at scales around 100 GeV, $R\simeq 0.5$ is an
  appropriate choice, clustering at the TeV scale requires $R\simeq 1$
  for optimal efficiency. We finally show that our results are valid
  in the presence of pileup, provided that a subtraction procedure is
  applied.
\end{abstract}

\maketitle

\section{Introduction}\label{sec:intro}

In collider physics, as soon as the final state involves hadronic
particles, jets become fundamental objects present in many
studies. Even if the idea of a jet as a bunch of collimated partons or
hadrons is what one always keeps in mind, the concept of a parton
itself is ambiguous and as a consequence different jet definitions
exist, possibly leading to different sets of jets for the same
hadronic event.

In practice, many jet definitions, which we shall review in Section
\ref{sec:jetdefs}, have been used for recent analysis. It turns out
however that, among them, some fail to satisfy the fundamental
requirements agreed upon in 1990 (Section \ref{sec:snowmass}). 

The first series of results presented in these proceedings addresses
those failures and give solutions that have been proposed
recently. This includes the \fastjet implementation of the $k_t$
algorithm, Section \ref{sec:ktspeed}, as well as the the SISCone and
anti-$k_t$ algorithms, Sections \ref{sec:siscone} and
\ref{sec:antikt}. Finally, in Section \ref{sec:filtering}, we will
briefly discuss a filtering technique using jet substructure that has
been recently proposed to reduce sensitivity to the underlying event.

Since one has the choice between different jet definitions to perform
jet analysis, a legitimate question is which of them is best suited
for a given analysis one wants to perform. In Section
\ref{sec:quality} of these proceedings, we answer that question in the
case of simple kinematic reconstructions at the LHC. To quantify the
efficiency of a jet definition, we introduce a figure of merit that is
directly related to an effective luminosity ratio. We will see that
not being careful enough in the choice of the jet algorithms and its
parameters (basically the radius $R$) can lead to important
consequences for potential discoveries at the LHC.

\section{Meeting fundamental requirements}\label{sec:jetdefs}

\subsection{Defining jets in the 20$^{\rm th}$  century}
\label{sec:pastdefs}

To begin with, let us briefly review the algorithms that have been
widely used over the past two decades for jet reconstruction in $pp$
collisions. Generally speaking, they fall in two categories that we
discuss hereafter.

{\bf Successive recombinations.} The first family of jet clustering
algorithms works by defining a distance between any pair of objects
and a beam distance for every object. One identifies the smallest
distance; if it is a beam distance, the object is called a jet and
removed from the event, otherwise, the two objects are recombined in a
single one. The procedure is repeated until no object are left in the
event. Two well-known examples of recombination algorithms are the
$k_t$ \cite{kt1,kt2} and Cambridge/Aachen (C/A) \cite{cam} algorithms,
using the distance
\begin{eqnarray}\label{eq:distance}
d_{ij} & = & \min(k_{t,i}^{2p},k_{t,j}^{2p})
             \left(\Delta y_{ij}^2 + \Delta \phi_{ij}^2\right),
             \nonumber\\[-2mm]
       &   & \\[-2mm]
d_{iB} & = & R^2\,k_{t,i}^{2p},\nonumber
\end{eqnarray}
where $p=1$ ($p=0$) corresponds to the $k_t$ (C/A) case.

{\bf Cone.} The cone algorithms aim at defining jets as
dominant directions of energy flow. To achieve that goal, one defines
the concept of {\em stable cone} as a circle of fixed radius in the
$(y,\phi)$ plane such that the sum of the 4-momenta of the particles
inside it points in the direction of its centre. 
Most of the cone algorithms used so far are {\em seeded} in the sense
that the search for stable cones starts from a given set of seeds from
which one iterates the cone contents until it is stable.
Since stable cones might overlap, one cannot define them as jets
directly. Two different techniques are in use to overcome this
overlapping problem.

The first option, that we shall refer to as {\em cone algorithms with
  split--merge}, first identify the set of all stable cones, then run
a split--merge procedure on them. The latter repeatedly identifies the
two hardest (originally, in $E_t$) overlapping cones; if their overlap
passes a threshold fraction (the overlap parameter $f$) of the softer
of the selected cones, they are merged. Otherwise, they are split by
associating every particle to the cone to whose centre it is closer.

For the seeded versions of the cone algorithms with split--merge, one
usually starts with all the particles in the event (usually with a
$p_t$ threshold) as seeds, as is the case for the CDF JetClu
\cite{jetclu} and ATLAS Cone algorithms. A step forward is to add as
new seeds the midpoints between all pairs of stable cones found after
this first pass. This is the case for the majority of the
recently-used cone algorithms \cite{Blazey}, noticeably the CDF
MidPoint, D0 run II Cone and PxCone algorithms.

The second solution to the problem of overlapping stable cones, {\em
  iterative cone algorithms with progressive removal (IC-PR)}, starts
by iterating the stable cone search from the hardest seed in the
event. The resulting stable cone is called a jet and its contents are
removed from the event. One then proceeds by iterating from the
hardest remaining seed, until all the particles are clustered. The
characteristic feature of this type of cone algorithm is that it
produces hard jets that are circular and soft-resilient. In other
words, the addition of soft particles does not modify the shape of the
hard jets, which is sometimes seen as an advantage for calibrating the
jets. The CMS Iterative Cone algorithm \cite{CMSit} falls in this
sub-category.

\subsection{Fundamental requirements}\label{sec:snowmass}

Back in 1990, a list of fundamental requirements that every jet
definition has to fulfil was agreed upon \cite{snowmass}. It is known
as the {\em SNOWMASS accords} and consists of the 5 following criteria:
\begin{compactenum}
\item simple to implement in an experimental
  analysis;\label{it:snow1}
\item simple to implement in the theoretical
  calculations;\label{it:snow2}
\item defined at any order of perturbation theory;\label{it:snow3}
\item yields finite cross section at any order of perturbation
  theory;\label{it:snow4}
\item yields a cross section that is relatively insensitive to
  hadronisation.\label{it:snow5}
\end{compactenum}

As consequences, because of constraints \ref{it:snow1} and
\ref{it:snow5}, we want the implementation of the algorithm to be fast
enough and as insensitive as possible to the underlying event (UE) so
that it can be used in experimental analysis. Also, the requirement
that the cross section remains finite at any order of perturbation
theory implies that the algorithm has to be infrared and collinear
(IRC) safe. Indeed, if it was not the case, cancellation between real
emissions and virtual corrections would not happen properly, leading
to divergences in perturbative cross-sections.

\subsection{Speeding up the $k_t$ algorithm}\label{sec:ktspeed}

\begin{figure}[ht]
\includegraphics[width=0.42\textwidth]{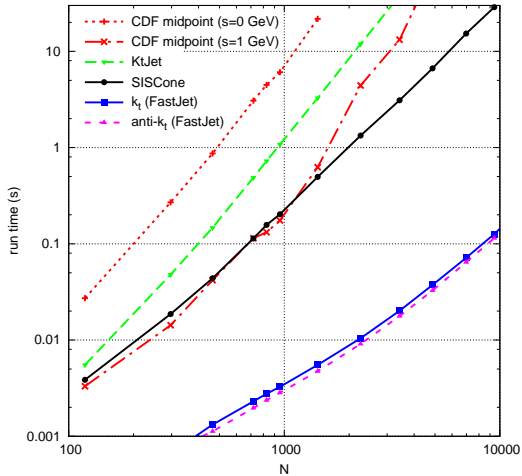}
\caption{Clustering time for various algorithms.}\label{fig:speed}
\end{figure}

Beside its rather large sensitivity to the UE, one argument that was
sometimes used against the $k_t$ algorithm was its relatively slow
running time. The {\em KtJet} \cite{ktjet} and {\em KtClus} \cite{kt1}
implementations, of complexity $\order{N^3}$ where $N$ is the number
of particles in the event, have a clustering time around 1 second for
$N=1000$, compared to about 0.2 seconds for the MidPoint cone with a 1
GeV seed threshold.

Remarkably, the complexity can be reduced to $\order{N \log(N)}$ using
computational-geometry techniques\footnote{For $N\lesssim 5000$, there
  exists a $\order{N^2}$ implementation that turns out to be faster
  than the $N\log(N)$ \cite{fastkt}.}  \cite{fastkt}. As seen on
figure \ref{fig:speed} ($k_t$ (\fastjet) curve), this leads to a
considerable improvement.

\subsection{SISCone and the IR unsafety of the MidPoint
  algorithm}\label{sec:siscone}

\begin{figure}[ht]
  \centering{\includegraphics[width=0.33\textwidth]{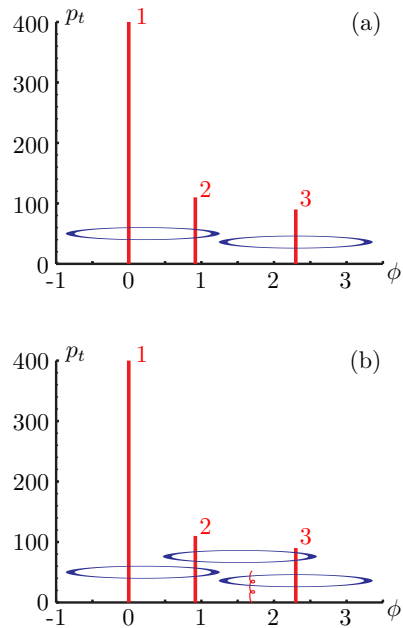}}
  \caption{Stable cones found by the MidPoint algorithm for (a) a
    3-particle event, (b) the same event with an additional infinitely
    soft gluon.}\label{fig:irfailure}
\end{figure}

It had already been noticed that the JetClu algorithm had some IR
unsafety problems {\em e.g.} when two hard particles were distant by
more that $R$ and less that $2R$, an additional infinitely soft gluon
added between the particles could change the clustering from 2 jets to
1 jet, yielding unreliable cross-sections at NLO in the inclusive jet
cross-section. The MidPoint algorithm was then introduced to cure that
problem.

Unfortunately, this is not the end of the story. Even if situations
with 2 hard particles in a common vicinity (plus another one, hadronic
or electroweak to balance $p_t$) are IR safe provided one uses the
MidPoint algorithm instead of JetClu, the problem has just been
shifted to situations with 3 hard particles in a common vicinity. This
is illustrated in figure \ref{fig:irfailure}, where we have clustered
a 3-hard-particle event twice --- with, fig. \ref{fig:irfailure}(b),
and without, fig. \ref{fig:irfailure}(a) an additional infinitely soft
gluon --- and notice that one finds different stable cones in the two
cases. This means that the MidPoint algorithm is also IR unsafe,
though one order further in the perturbative expansion in the strong
coupling than the JetClu algorithm.

To solve that problem to all orders in the perturbative expansion, we
first notice that the IR unsafety comes from the fact that, in the
event without the soft gluon, the stable cone enclosing particles 2
and 3 has been missed. Since the mathematically well-defined set of
stable cones is IRC safe, {\em i.e.} it changes neither when splitting
a particle collinearily nor when adding infinitely soft particles (up
to harmless stable cones made only of those soft particles), finding a
method that provably identifies all stable cones, guarantees IRC
safety\footnote{We also need to be careful about the IRC safety of the
  split--merge procedure. Actually, a couple of technical issues, such
  as the choice of the variable used to order the overlapping cones,
  have to be dealt with. We will not address them here (see
  \cite{siscone} for details).}.

We then observe that every circular enclosure of given radius $R$ in
the $(y,\phi)$ plane can be translated in any direction until it
touches one point, then rotated around that point until it touches a
second one, without changing its contents. Therefore, enumerating all
pairs of points, and for every pair of points considering the two
circles of radius $R$ they define and the four possible
inclusion/exclusion states of the edge particles, we enumerate all
possible enclosures. For each of them we can then test if it is
stable or not, which solves our problem.
The complexity of this algorithm is $\order{N^3}$ (a factor $N^2$
coming from the enumeration of the pairs of parent points and an
additional factor of $N$ to test the stability of every enclosure). It
is actually possible, using extra geometric observations, to improve
that complexity to $\order{N^2 \log(N)}$. This has been implemented
\cite{siscone} in a new algorithm named {\em SISCone} (Seedless
Infrared Safe Cone).
Fig. \ref{fig:speed} shows that SISCone runs faster than the
IR-unsafe MidPoint algorithm (of order $N^3$), even when a seed
threshold of 1 GeV is applied.

\subsection{Anti-$k_t$ and the collinear unsafety of the IC-PR
  algorithm}\label{sec:antikt}

\begin{figure}[ht]
  \centering{\includegraphics[width=0.33\textwidth]{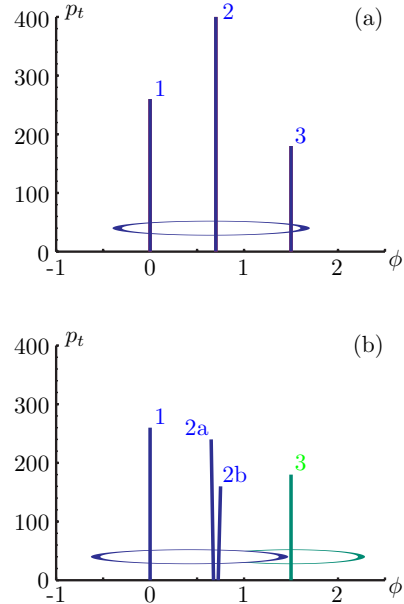}}
  \caption{Jets found by the iterative cone for (a) a 3-particle event,
    (b) the same event with a collinear splitting.}\label{fig:uvfailure}
\end{figure}

In a similar way as for the case of the MidPoint algorithm, we can
show that the iterative cone algorithm with progressive removal
(IC-PR) also suffers from divergences in the perturbative series, at
the same order as MidPoint, this time due to collinear unsafety.
In order to see that, first consider the event of fig.
\ref{fig:uvfailure}(a). Iterating from the hardest seed gives one jet
containing all particles. If one splits the hardest of these particles
in two collinear ones, fig. \ref{fig:uvfailure}(b), iteration
starts with the leftmost particle and 2 jets are found. This means
that the IC-PR is collinear unsafe at the level of 3
particles (+1 to balance $p_t$).

To address this issue, we will go back to the recombination-type
algorithms. We have already mentioned that setting $p=0$ or $1$ in
eq. (\ref{eq:distance}) reproduces the C/A and $k_t$ algorithms
respectively. We now introduce the {\em anti-$k_t$ algorithm},
corresponding to $p=-1$ in eq. (\ref{eq:distance}) \cite{antikt}.

At first sight, it is not obvious what this new, IRC-safe, algorithm
has to do with the IC-PR. However, since hard particles will be
associated a small anti-$k_t$ distance, they will grow in circles,
clustering softer particles in their vicinity up to a distance
$R$. This thus leads to soft-resilient hard jets, {\em i.e.}  the
boundary of the hard jets is not affected by soft radiation, the
precise characteristic feature of the IC-PR.

Finally, the anti-$k_t$ algorithm is also amenable to a fast
implementation (see fig. \ref{fig:speed}) using the same techniques as
for the $k_t$ algorithm (see Section \ref{sec:ktspeed}).

\subsection{Filtering}\label{sec:filtering}

One of the promising potential improvement of jet clustering is to
make use of the jet substructure. For example, we can use the
following filtering technique to reduce the contamination due to the
underlying event:
\begin{compactenum}
\item Cluster the event using, {\em e.g.}, the C/A algorithm with a
  radius $R$,
\item For each jet, recluster it using a smaller radius
  $R_{\text{sub}}$ and keep only the $n_{\text{sub}}$ hardest jets as
  part of the initial jet, throwing the other subjets.
\end{compactenum}
The aim of filtering is to remove contamination due to soft background
like the underlying event while keeping as much as possible of the
perturbative radiation.
This has already proven \cite{filtering} to be efficient in Higgs
searches in the $b\bar b$ channel.

In what follows we shall use $R_{\text{sub}}=R/2$ and
$n_{\text{sub}}=2$, though a more extensive study of the effects of
these parameters would be interesting.

\section{Quantifying kinematic reconstruction efficiency}\label{sec:quality}

Now that we dispose of 5 IRC-safe jet algorithms --- the $k_t$, C/A,
anti-$k_t$, SISCone and C/A+filtering algorithms, all available from
\fastjet \cite{fastjet} --- we may ask, given an analysis involving
jets we want to perform, which jet definition, {\em i.e.} the jet
algorithm and its parameters, is best suited.
In this Section we address that question for the case of kinematic
reconstructions at the LHC. 

We first introduce a series of benchmark processes we will
investigate, then a figure of merit that allows one to quantify the
performance of a jet definition and finally present our results, both
with and without including pileup. For a more extensive discussion,
see \cite{lh07sm,jetalgs}.

\subsection{Benchmark processes}\label{sec:processes}

We will study the following 3 processes:
\begin{compactitem}
\item $Z'\to q\bar q$ as a source of quark jets. We reconstruct the
  $Z'$ from the 2 hardest jets in the event (imposing a maximal
  rapidity difference $|\Delta y|\le 2$ between them).
\item $H\to gg$ as a source of gluon jets. We reconstruct the $H$
  from the 2 hardest jets in the event (imposing again $|\Delta y|\le
  2$).
\item a $t\bar t$ pair with fully hadronic decay into 6 jets (4 jets
  from the 2 $W$ bosons and 2 jets from the $b$ and $\bar b$
  quarks). The two $W$'s and then the 2 tops are reconstructed from
  the 6 hardest jets in the event\footnote{Practically, $b$-jets are
    tagged assuming the $b$ mesons are stable; then the two $W$ are
    reconstructed by pairing the 4 remaining jets so as to minimise
    $(M_{i_1i_2}-M_W)^2+(M_{i_3i_4}-M_W)^2$; finally, the two top jets
    are reconstructed by matching the $b$-jets with the $W$ so as to
    minimise the mass difference between the 2 top candidates.}.
\end{compactitem}
For the case of quark and gluon jets, we can study the scale
dependence by varying the mass of the $Z'/H$ boson (in practice,
between 100 GeV and 4 TeV).
The last case, fully hadronic $t\bar t$ decay, allows us to study the
relevance of our results in more complex environments, where one risks
tensions between resolving the jets and capturing the perturbative
radiation.
In each of these situations we have generated a sample of events using
Pythia 6.4 tune DWT.

\subsection{Figure of merit}\label{sec:measure}

\begin{figure*}
  \centerline{
    \includegraphics[width=0.42\textwidth]{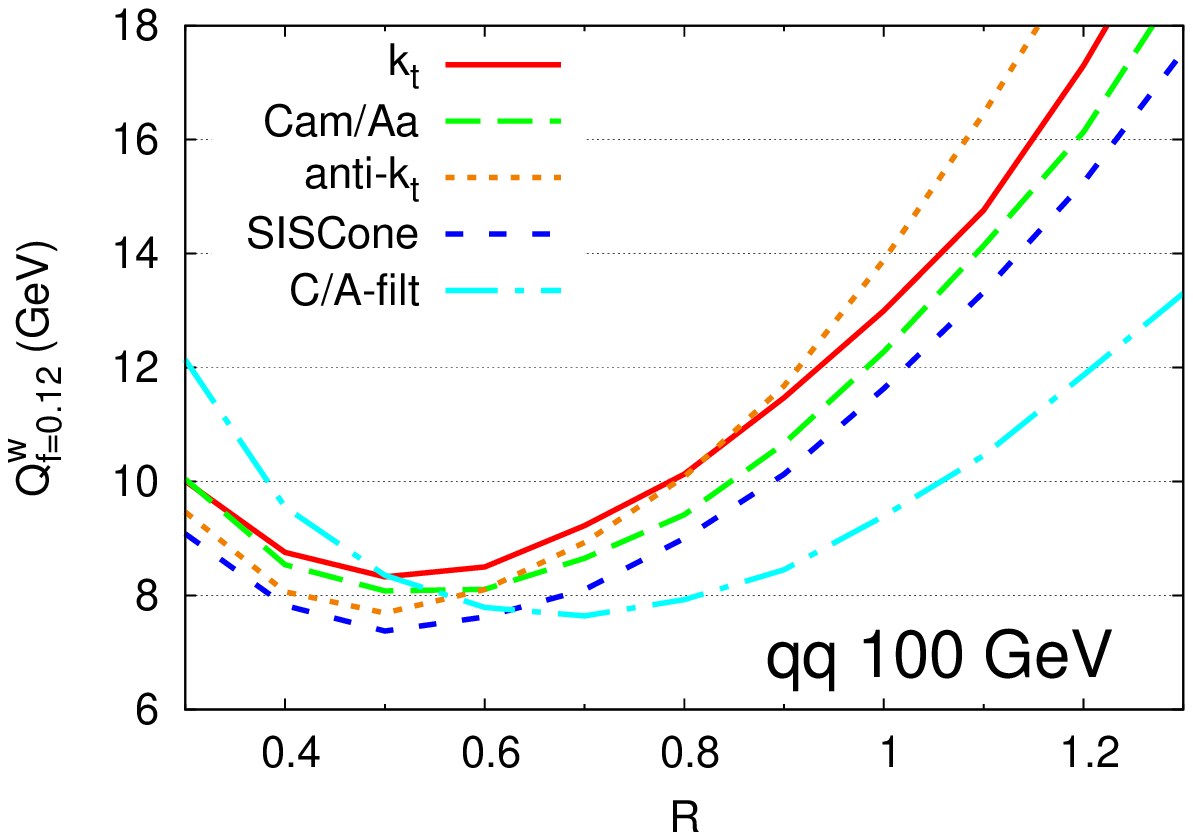}
    \includegraphics[width=0.42\textwidth]{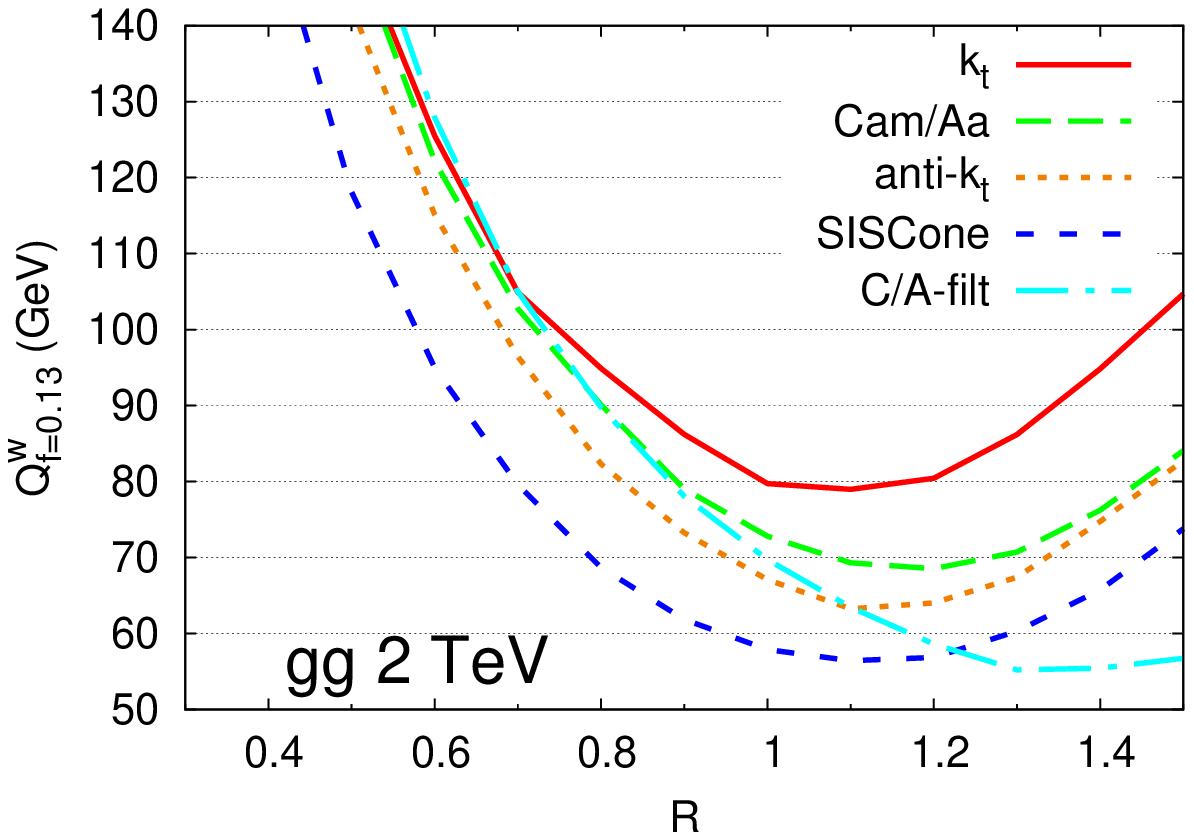}
  }
  \caption{result for the quality measure for quark jets at 100 GeV
    (left) and gluon jets at 2 TeV (right). The results are displayed
    as a function of $R$ and each curve corresponds to a given
    algorithm.}\label{fig:qa}
\end{figure*}

\begin{figure*}
  \centerline{\includegraphics[width=0.8\textwidth]{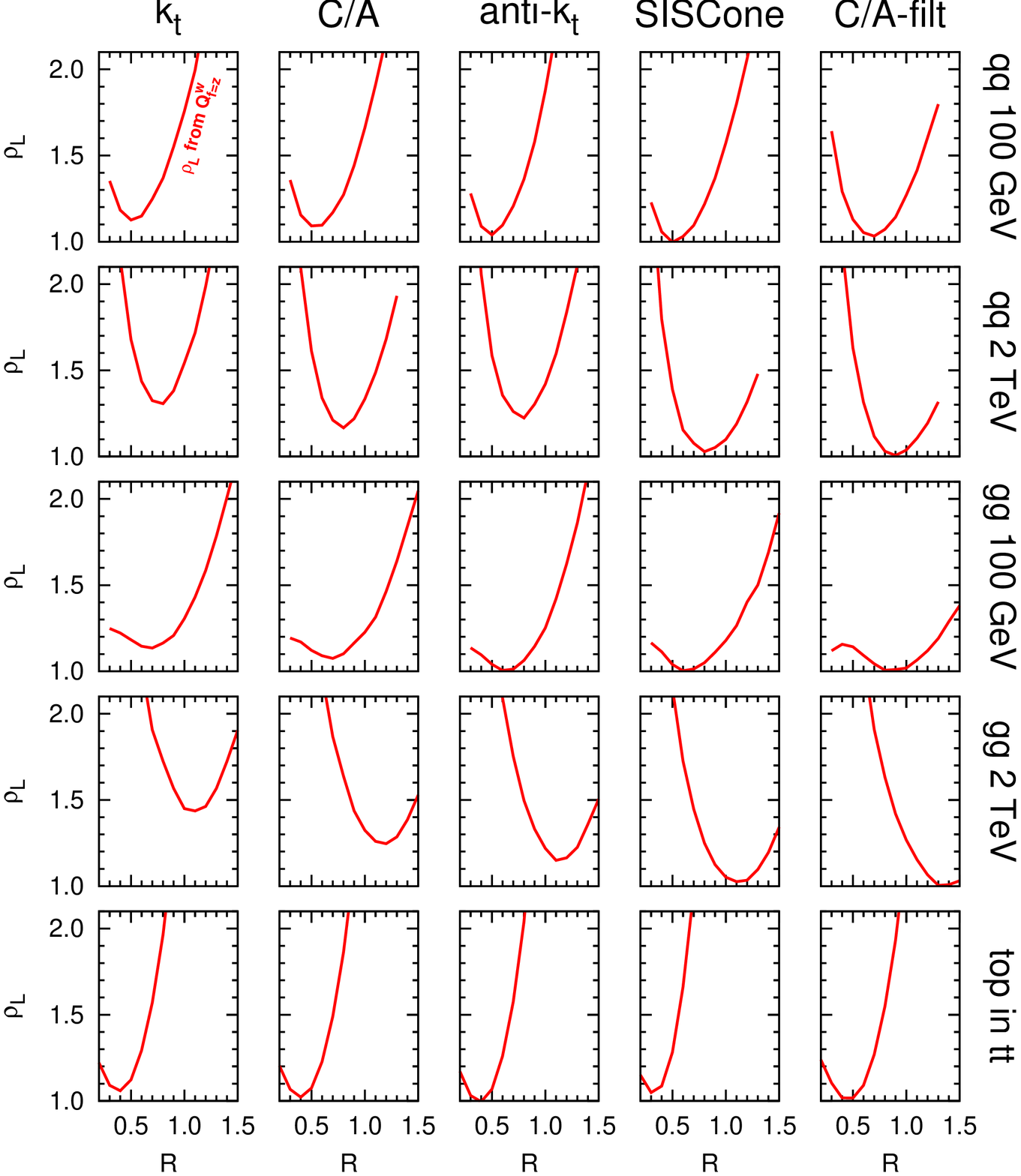}}
  \caption{Effective luminosity ratios as a function of the parameter
    $R$. Each line corresponds to a different process and each
    column to a different algorithm. For each process, $\rhol$ is
    normalised to the best possible definition for that
    process.}\label{fig:rhol}
\end{figure*}

In order to quantify which jet definition is performing better than
another, we need a figure of merit. 
Since jets are expected to represent an original parton, one might be
tempted to quantify the efficiency by comparing the jets to the
initial partons. But because partons are a ill-defined concept
(especially at NLO), this is not robust enough.

Another option would be to fit a given distribution, {\em e.g.} a
Gaussian, to the peak in the mass spectrum. But since the shape of the
peak can be asymmetric, this would not produce a reliable result
either. 

For these reasons, we will quantify the efficiency of a jet definition
based on the observation that, if two peaks have similar number of
events then the narrower is the better one. We therefore introduce the
quality measure \Qa{z} as the {\em width of the smallest possible
  window that contains a fraction $f=z$ of the events}. The smallest
values of \Qa{z} should then correspond to the most efficient jet
definitions.

This intuitively does what we want in the sense that a ``better'' jet
definition should contain a given fraction of the events in a smaller
window and therefore have a smaller \Qa{z}.

This quality measure can be related to a variation of luminosity
needed to maintain a constant significance for a signal relative to
background. For a jet definition $\JD$, the latter is defined as
$\Sigma(\JD) \equiv N_{\rm signal}^\JD/ \sqrt{N^\JD_{\rm
    bkgd}}$. Assuming a constant background, we thus have
\[
\frac{\Sigma(\JD_1)} {\Sigma(\JD_2)}
  = \left[\frac{N^{\JD_2}_{\rm bkgd}}{N^{\JD_1}_{\rm bkgd}}\right]^{1/2}
  = \left[\frac{Q_{f=z}^{w}(\JD_2)} {Q_{f=z}^{w}(\JD_1)}\right]^{1/2}.
\]
two jet definitions $\JD_1$ and $\JD_2$. This relation comes from the
fact that the number of signal events is fixed by the fraction $f=z$
in the quality measure, while the background is directly proportional
to the width of the window {\em i.e.} to \Qa{z}. A ``better''
definition, {\em i.e.} a smaller \Qa{z}, thus corresponds to a larger
discriminating power.

We can then define an effective luminosity ratio
\begin{eqnarray}\label{eq:rhol}
  \rho_{\cal L}(\JD_2 / \JD_1) 
    & \equiv &
      \frac{{\cal L}(\text{needed with }\JD_2)}
           {{\cal L}(\text{needed with }\JD_1)}
           \\
    & = & \left[ \frac{\Sigma(\JD_1)}{\Sigma(\JD_2)} \right]^2
        = \frac{\Qa{z}(\JD_2)}{\Qa{z}(\JD_1)}.\nonumber
\end{eqnarray}
This means that a jet definition $\JD_1$ with a quality measure twice
as large as $\JD_2$ will need twice the integrated luminosity in order
to achieve the same discriminating power as $\JD_2$.

\subsection{Results without pileup}\label{sec:resnopu}

For the processes presented in Section \ref{sec:processes}, we have
clustered our event samples using the 5 IRC safe jet algorithms and
varying the parameter\footnote{In the case of SISCone, the overlap
  threshold $f$ has been fixed to 0.75, the preferred choice at the
  time being.} $R$ between 0.1 and 1.5. We can then compute the
quality measure in each of those cases (see figure \ref{fig:qa}),
which allows us to (i) compare different algorithms, and (ii), for a
given algorithm, find the optimal value $R_{\text{best}}$ of the
parameter $R$.

Figure \ref{fig:rhol} summarises our results in a more compact and
physical way: we have plotted the effective luminosity ratios
corresponding to a series of selected processes: quark and gluon jets
at 100 GeV and 2 TeV, and top reconstruction in the $t\bar t$
sample. For every process, \rhol has been normalised to the best jet
definition for that process.

We observe a few important features: first of all, in general, SISCone
and C/A with filtering tend to perform slightly better than the $k_t$,
C/A and anti-$k_t$ algorithms, especially at higher scales (with the
exception of the top reconstruction where all algorithms have similar
performances). For example, for the 2 TeV gluon case, choosing the
$k_t$ algorithm instead of SISCone or C/A with filtering which are the
preferred choices, translates into a cost of nearly 50\% in the
effective luminosity ratio.

Furthermore, the efficiency of a jet definition strongly depends on
$R$ and not choosing the preferred value $R_{\text{best}}$ can be very
costly. On top of that, $R_{\text{best}}$ varies significantly from
one process to another. It increases from $R_{\text{best}}\simeq 0.5$
at small scales up to $R_{\text{best}}\simeq 1$ at TeV
scales. $R_{\text{best}}$ is also bigger for gluon jets than for quark
jets. Practically, using SISCone with $R=0.5$, the preferred choice
for quark jets at 100 GeV, to cluster gluon jets at 2 TeV costs a
factor of 2 in \rhol compared to the best definition for that
process. Conversely, using the preferred value for gluon jets at 2 TeV
--- {\em i.e.} SISCone with $R=1.1$ or C/A with filtering and $R=1.3$
--- to cluster quark jets at 100 GeV also leads to \rhol increasing by
at least 50\%.

\subsection{Results with pileup}\label{sec:reswpu}

\begin{figure*}
  \centerline{\includegraphics[width=0.8\textwidth]
    {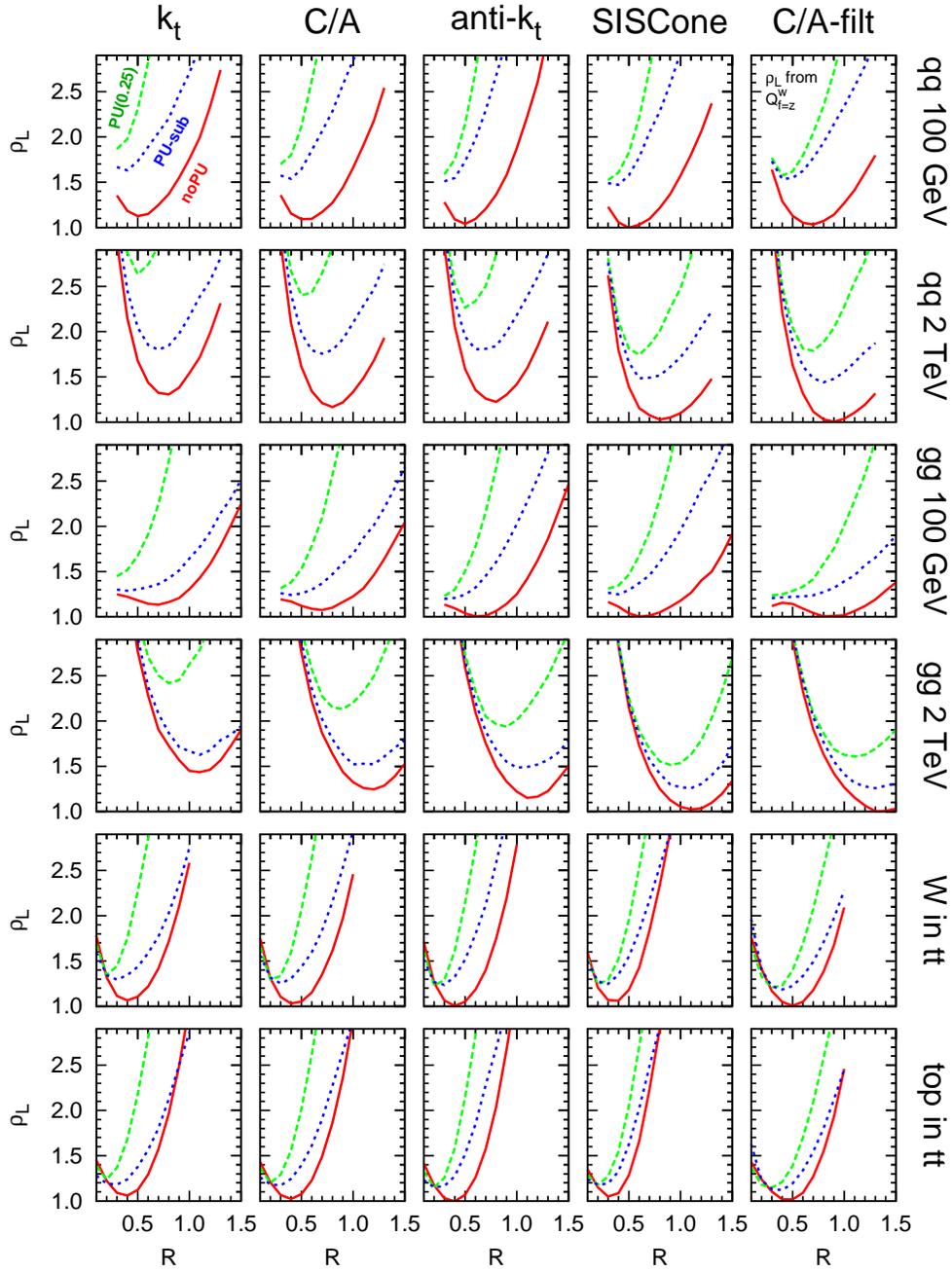}}
  \caption{Same as for figure \ref{fig:rhol}, now including the
    effects of pileup. The solid (red) curve corresponds to the
    situation without pileup, the dashed (green) one to the case of
    unsubtracted high-luminosity pileup and the dotted (blue) curve to
    the situation where pileup is subtracted.}\label{fig:rholpu}
\end{figure*}

Pileup corresponds to the fact that multiple $pp$ interactions can
happen at the same time. For the case of the LHC at the designed
luminosity, one typically has an average of about 25 collisions per
bunch crossing.
This produces a large number of additional soft particles that form a
reasonably uniform background over the detector. 

For our concerns, pileup has two consequences. Because it adds a
background to every jets, the $p_t$ of the jets will be overestimated
and the position of the mass peak in our kinematic reconstructions
will be shifted towards larger masses. Also, because the amount of
pileup varies from one event to another, the reconstructed peak will
be smeared, an effect that directly affects our quality measure.

We therefore want a (simple and generic enough) method to subtract the
contamination from the pileup background on an event-by-event
basis. In practice, we will follow the subtraction method suggested in 
\cite{subtraction}: for each jet $j$ in an event, we first compute its 
(4-vector) area \cite{areas} $A_j^\mu$. The subtracted jet is then
obtained using
\begin{equation}
p_{j,{\rm sub}}^\mu = p_j^\mu - \rho\,A_j^\mu,
\end{equation}
where $\rho$ is the average density of pileup per unit area. Since the
density of background is reasonably uniform, we estimate $\rho$ for
each event using \cite{subtraction} $\rho =
\text{median}\{p_{t,j}/A_{t,j}\}$, where all the jets (up to a maximal
rapidity) are included in the computation of the median.

In practice, we have considered the same benchmark processes as in the
study without pileup. Pileup is added to each event under the form of
a random (Poisonian) number of minimum bias events, also generated
with Pythia 6.4 tune DWT. We apply the same reconstruction procedure
as above and compute the quality measure \Qa{z} with pileup effects
subtracted or not.

We illustrate our results by showing on figure \ref{fig:rholpu} the
effective luminosity ratio \rhol as obtained for a few representative
processes. We compare the results with pileup (subtracted or not) with
the corresponding quality measures obtained without pileup (solid
curves).

The first observation is that if the pileup is not subtracted (dashed
curves), it causes a large degradation of the quality measure. The
preferred $R$ is also shifted to smaller values, which one can explain
by the fact that the contamination due to pileup background is smaller
at small $R$.
The second element of information comes from the case where we apply
our subtraction method (dotted curves). Obviously, even if the
subtraction is not perfect --- the quality is still larger than before
pileup addition --- one sees a significant improvement compared to the
situation without subtraction. This means that the subtraction method
gives narrower peaks {\em i.e.} the smearing due to pileup
fluctuations between events is strongly reduced by the subtraction.
On top of that, after the subtraction has been performed, the preferred
value for $R$ obtained in the situation without pileup is no longer
strongly disfavoured to the profit of a smaller $R$. This is important
since it implies that our conclusions from Section \ref{sec:resnopu}
are still valid in the presence of pileup provided one uses
subtraction.

\section{Conclusions}\label{sec:ccl}

\begin{table}[ht]
  \begin{tabular}{|l|l|c|l|}
    \hline
    Algorithm & Type  & IRC \\
    \hline
    \hline
    inclusive $k_t$  \cite{kt1,kt2} & SR$_{p=1}$   & OK
    \\ \hline
    Cambridge/Aachen \cite{cam}     & SR$_{p=0}$   & OK
    \\ \hline
    anti-$k_t$ \cite{antikt}        & SR$_{p=-1}$  & OK
    \\ \hline
    SISCone \cite{siscone}          & SC-SM        & OK
    \\ \hline
    CDF JetClu \cite{jetclu}        & IC$_r$-SM    & IR$_{2+1}$
    \\ \hline
    CDF MidPoint \cite{Blazey}      & IC$_{mp}$-SM & IR$_{3+1}$ 
    \\ \hline
    D0 Run II cone \cite{Blazey}    & IC$_{mp}$-SM & IR$_{3+1}$
    \\ \hline
    ATLAS Cone                      & IC-SM        & IR$_{2+1}$
    \\ \hline
    CMS Iterative Cone \cite{CMSit} & IC-PR        & Coll$_{3+1}$
    \\ \hline
  \end{tabular}

  \caption{\small Overview of some jet algorithms used recently 
    in experimental or theoretical work. 
    SR$_{p=x}=$ sequential recombination (with $p=0,\pm 1$, 
    see (\ref{eq:distance})); SC = seedless cone (finds all cones); 
    IC = iterative cone (with midpoints $mp$, ratcheting $r$), using 
    either split--merge (SM) or progressive removal (PR) in order to
    deal with overlapping stable cones.
    Regarding IRC status, given $n$ hard particles in a common
    neighbourhood, IR$_{n+1}$ indicates that the addition of 1 extra
    soft particle can modify the number of final hard jets, while  
    Coll$_{n+1}$ means that the collinear splitting of one of the
    particles can modify the number of final hard jets.
  }\label{tab:algs}
\end{table}

The first point we have addressed concerns the jet algorithms
themselves. As summarised in table \ref{tab:algs}, some of the
commonly used algorithms fail to satisfy the fundamental requirements
of infrared and collinear safety. We have introduced SISCone and the
anti-$k_t$ algorithm to cure infrared unsafeties of the MidPoint
algorithm and collinear unsafeties of the IC-PR. 
We have also mentioned a new implementation of the $k_t$ algorithm
leading to a sizeable improvement in speed.

We have then addressed the problem of quantifying the performances of
the jet definitions in the case of kinematic reconstructions at the
LHC. From that study, one learns important messages showing that
optimisation of the jet definition can significantly improve the
potential for an early discovery at the LHC.
We have shown that the preferred jet definition varies significantly
from one process to another. SISCone and C/A with filtering tend to
perform slightly better than the other algorithms and the preferred
value for $R$ goes from $R\simeq 0.5$ for 100 GeV jets to $R\simeq 1$
for TeV jets with a slightly larger value for gluon jets compared to
quark jets. This indicates that a single choice for $R$ is not
sufficient to cover the whole kinematic range at the LHC.

Finally, if one includes the effects of pileup, we have seen that,
provided we use an adequate subtraction procedure, the conclusions
obtained in the analysis without pileup hold once pileup is added,
noticeably, there is no need to take a smaller value of $R$.

In the more complex case of top reconstruction where jet clustering
involves tension between catching enough radiation to reconstruct the
jet energy, and resolving the 3 jets coming from the top decay
products, techniques like the subjet analyses might be of crucial
importance and definitely deserve further studies.

\end{document}